# Lecture notes on Machine Learning applications for global fits

**Jorge Alda**[1,2⋆]

**1** Dipartimento di Fisica e Astronomia "Galileo Galilei", Università degli Studi di Padova and INFN Sezione Padova, via Marzolo 8 35129 Padova, Italy.

**2** Centro de Astropartículas y Física de Altas Energías (CAPA), Pedro Cerbuna 12 50009 Zaragoza, Spain.

⋆ jorge.alda@pd.infn.it

## Abstract

**These lecture notes provide a comprehensive framework for performing global statistical fits in high-energy physics using modern Machine Learning (ML) surrogates. We begin by reviewing the statistical foundations of model building, including the likelihood function, Wilks' theorem, and profile likelihoods. Recognizing that the computational cost of evaluating model predictions often renders traditional minimization prohibitive, we introduce Boosted Decision Trees to approximate the log-likelihood function.**
**The notes detail a robust ML workflow including efficient generation of training data with active learning and Gaussian processes, hyperparameter optimization, model compilation for speed-up, and interpretability through `SHAP` values to decode the influence of model parameters and interactions between parameters. We further discuss posterior distribution sampling using Markov Chain Monte Carlo (MCMC).**
**These techniques are finally applied to the $B^\pm \to K^\pm \nu\bar{\nu}$ anomaly at Belle II, demonstrating how a two-stage ML model can efficiently explore the parameter space of Axion-Like Particles (ALPs) while satisfying stringent experimental constraints on decay lengths and flavor-violating couplings.**

## Contents

# 1 Introduction

These lecture notes cover the talk and hands-on tutorial "Machine Learning for global fits" of the 4th COMCHA School on Computing Challenges, celebrated in Zaragoza (Spain) from 8 to 15 April 2026. The methods explained in these notes draw mainly from my experience in applying Machine Learning to global fits for flavour physics in the Standard Model Effective Field Theory [1–3], updated with some of the latest developments.

The lecture notes are structured as follows: in Sec. 1.1, I will motivate the pertinence of Machine Learning methods by reviewing the role of the likelihood function in global fits and the computational challenges it might pose. I will present the various Machine Learning techniques in Sec. 2: in Sec. 2.1 Gaussian processes for the generation of a training dataset using active learning, in Sec. 2.2 decision trees (and particularly XGBoost) for regression, in Sec. 2.3 SHAP values for model explainability, and in Sec. 2.4 the Markov chain Monte Carlo sampling. In Sec. 3, I will present the physical problem that we will tackle in the tutorial session with the tools previously explained: an excess in the decay rate of $B^\pm \to K^\pm \nu\bar{\nu}$ explained by a long-lived light particle called Axion-like particle. Finally, I will wrap up with conclusions in Sec. 4.

The repository Jorge-Alda/comcha_tutorial contains the code used in the tutorials.

## 1.1 The likelihood function

Our job as physicists is to build and test **models** that aim to describe the world around us. Models produce quantitative **predictions** for measurable quantities or **observables**, such as decay rates, cross sections or kinematical distributions. Models are given in terms of **parameters**, which in general we will denote as a vector $\theta \in \Theta$, where $\Theta$ is the parameter space. In general, $\theta$ will include both parameters of interest and nuisance parameters (backgrounds, detector parameters, etc.). Global fits try to find a consistent set of parameters that describes simultaneously every available observation.

The main statistical tool in model building is the likelihood function $\mathscr{L}(\theta)$, which is defined as the conditional probability of obtaining the observed experimental data given the values of the parameters $\theta$,

$$\mathscr{L}(\theta) = p(\mathrm{data}|\theta)\,. \tag{1}$$

I will often use the slight abuse of notation $\chi^2(\theta) \equiv -2\log\mathscr{L}(\theta)$. A simple but illustrative example of likelihood is the case of $N$ statistically independent observables with gaussian errors,

$$\mathscr{L}(\theta) = \prod_{i=1}^{N} \frac{1}{\sqrt{2\pi}\sigma_i} \exp\left(-\frac{(y_i - x_i(\theta))^2}{2\sigma_i^2}\right)\,, \tag{2}$$

where $y_i$ are the experimental data, $x_i(\theta)$ the predictions evaluated at the parameter point $\theta$, and $\sigma_i^2 = \sigma_{y_i}^2 + \sigma_{x_i}^2$ is the sum of experimental and theoretical variances. If we define $\chi_i(\theta) = \frac{y_i - x_i(\theta)}{\sigma_i}$, then $\chi^2(\theta) = \sum \left(\chi_i(\theta)\right)^2$ up to an irrelevant constant.

We can use the observed data to estimate the parameters of the model as Maximum Likelihood Estimators (MLE) $\hat{\theta}$,

$$\hat{\theta} = \arg\max \mathscr{L}(\theta) = \arg\min \chi^2(\theta). \tag{3}$$

In the specific case of the likelihood in Eq. (2), the MLE are equivalent to least squares fitting.

Another application of the likelihood function is hypothesis testing, using Wilks' theorem. Given a null hypothesis $\hat{\theta}_0 \in \Theta_0$ ("background-only") and an alternative hypothesis $\hat{\theta}_1 \in \Theta_1$ ("background + signal"), such that the null hypothesis must be nested $\Theta_0 \subset \Theta_1$ (and consequently the dimensionality $n_0$ of $\Theta_0$ is smaller than the dimensionality $n_1$ of $\Theta_1$) then the test statistic

$$\lambda = -2\log \frac{\mathscr{L}(\hat{\theta}_0)}{\mathscr{L}(\hat{\theta}_1)} = \chi^2(\hat{\theta}_0) - \chi^2(\hat{\theta}_1) \tag{4}$$

is a random variable whose distribution, under certain regularity conditions[1], asymptotically converges to a $\chi^2$-distribution with $\nu = n_1 - n_0$ degrees of freedom. Then the $p$-value is defined as

$$p = \int_\lambda^\infty f_{\chi^2}(t;\nu)\, dt, \tag{5}$$

where $f_{\chi^2}$ is the probability distribution function of the $\chi^2$ distribution. The $p$-value represents the probability of obtaining a value of the test statistics larger than $\lambda$ under the null hypothesis. In particle physics it is usual to instead report the significance $Z$, in units of $\sigma$, defined as

$$Z = \Phi^{-1}(1-p), \tag{6}$$

where $\Phi^{-1}$ is the inverse of the cumulative distribution function of the Gaussian.

We can also use Wilks' theorem to estimate the error for the MLE of the parameter $\theta^i$. We first define the profile likelihood $\mathscr{L}_P(\theta^i)$ (which despite its name, it is not a likelihood function) as

$$\mathscr{L}_P(\theta^i) = \max_{\theta^{j\neq i}} \mathscr{L}(\theta^1, \dots, \theta^n). \tag{7}$$

Then

$$\lambda_P(\theta^i) = -2\log \frac{\mathscr{L}_P(\theta^i)}{\mathscr{L}(\hat{\theta})} \tag{8}$$

is distributed according to a $\chi^2$ distribution with 1 degree of freedom. As such, the values of $\theta^i$ for which $\lambda_P(\theta^i) = 1$ define the $1\,\sigma$ lower and upper bounds (and analogously, $\lambda_P(\theta^i) = k^2$ defines the $k\,\sigma$ bounds). Similarly, for the estimation of the covariance of $\theta^i$ and $\theta^j$, using the profile likelihood $\mathscr{L}_P(\theta^i, \theta^j)$ and setting $\lambda_P(\theta^i, \theta^j) = 2.30$, one obtains a roughly elliptical contour, and the orientation of its axes is related to the covariance.

While this program might seem straightforward, it critically depends on our capability to evaluate the likelihood function in an efficient way, as numerical minimization and scanning typically requires hundreds or thousands of evaluations. In practice, we many times encounter likelihoods that depend on a large number of correlated, non-gaussian observables, and/or where the model predictions $x_i(\theta)$ are too computationally costly.

[1] One of the regularity conditions is that $\hat{\theta}_0$ must be in the interior of $\Theta_0$. If $\hat{\theta}_0$ lies in the boundary of $\Theta_0$, then the asymptotic distribution is a combination of $\chi^2$ and Dirac $\delta$. [4]

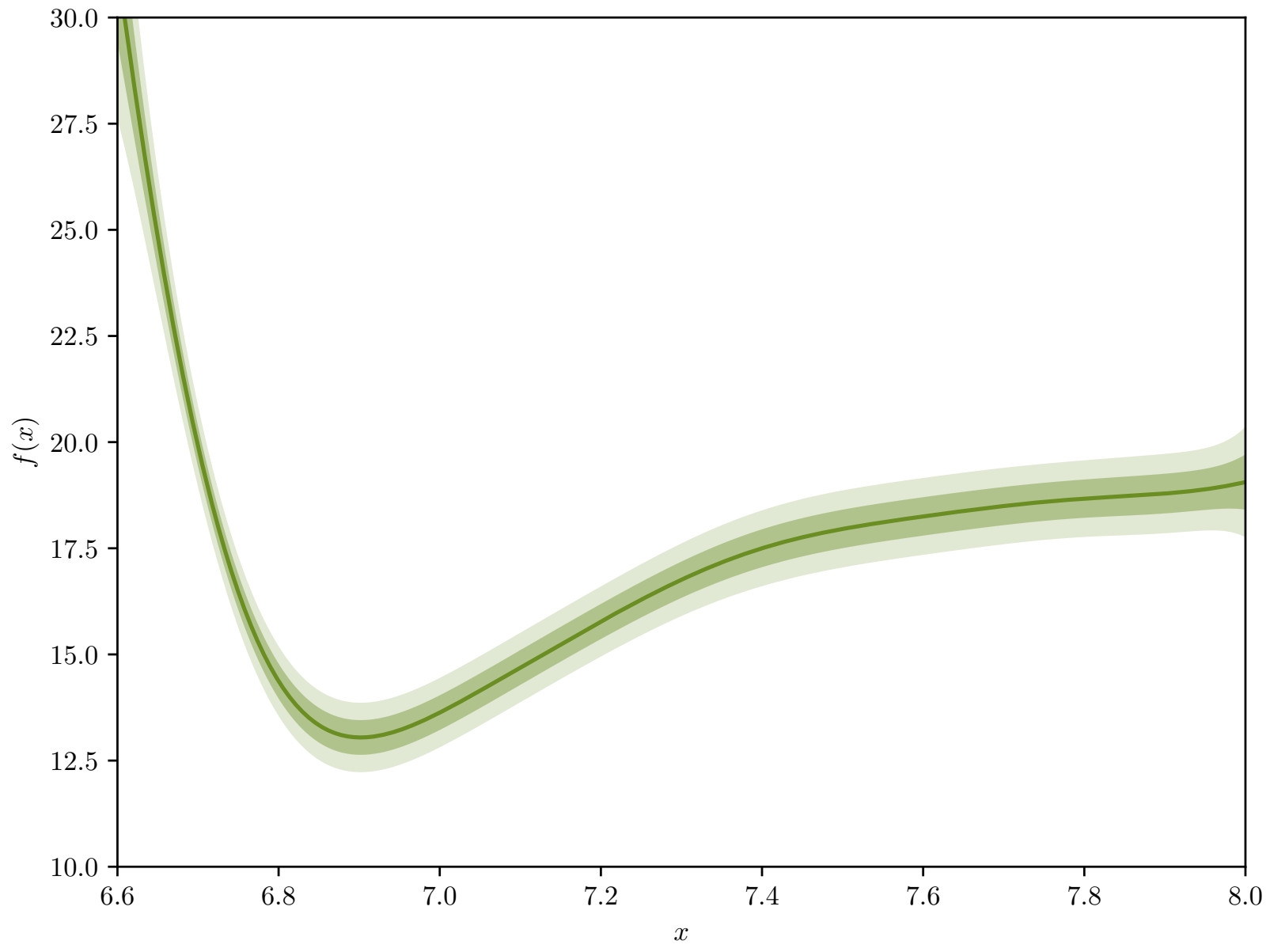

Figure 1: Example of the prediction of a trained Gaussian process: The line denotes the central value of the prediction for each $x$, and the shaded regions, the values allowed at $1\sigma$ and $2\sigma$.

## 2 Machine Learning techniques

In order to circumvent the limitations imposed by potentially expensive likelihood function evaluations, we are going to explore the possibility of training a Machine Learning surrogate model to approximate the log-likelihood and dramatically improve the computational efficiency.

### 2.1 Active learning

The first problem that we have to face is the creation of a training dataset that adequately captures the properties of the likelihood function, while keeping the number of evaluations to a minimum. An useful strategy is active learning: starting from a very small dataset (for example, drawn randomly from a latin hypercube), we will let a ML algorithm to recursively to decide which points to evaluate and add to the dataset, based on the trade-off two criteria [5]:

- **Exploitation:** Add new points near the optimum (largest likelihood, least $\chi^2$) according to the current knowledge of the learner.
- **Exploration:** Add new points in regions where the predictions of the learner are more uncertain, in order to increase the knowledge of the learner.

Therefore we need to use a ML algorithm that produces an estimation of the uncertainty, as well as the prediction itself. A natural choice are Gaussian processes (GP), the generalization of multivariate Gaussian probability distribution to infinite dimension. A GP is determined by the mean function $\mu(x)$ and covariance function $k(x,x')$, such that a random function is distributed according

$$f(x) \sim \mathcal{GP}(\mu(x), k(x,x')) \,. \tag{9}$$

The values of $f$ evaluated at a finite number of points are a random variable that follows a multivariate Gaussian distribution of the given mean and covariance, as exemplified in Fig 1.

Expected Improvement (EI) acquisition function for Gaussian Processes is a simple strategy that balances exploitation and exploration. Given a current best point $x^*$, the prediction improvement for a candidate point $x$ is

$$I(x) = \frac{s(f(x^*) - \mu(x)) - \delta}{\sigma(x)} \tag{10}$$

where $f(x)$ is the trained GP, $s = 1$ for minimization or $s = -1$ for maximization, and $\delta$ controls the exploration-exploitation trade-off ($\delta = 0$ is only exploitation, $\delta \sim |f(x^*)|$ is mostly exploration). Then the EI is given by

$$EI(x) = \sigma(x)[I(x)\Phi(I(x)) + \varphi(I(x))], \tag{11}$$

where $\Phi$ and $\varphi$ are respectively the CDF and PDF of a Gaussian distribution of mean 0 and variance 1.

$$\frac{\partial EI}{\partial \mu(x)} = \Phi(I), \qquad \frac{\partial EI}{\partial \sigma(x)} = \varphi(I), \tag{12}$$

EI is a monotonic function with respect to both the mean and the uncertainty. The maximum sensitivity to changes in the mean is achieved when $\mu \ll f(x^*) - \delta$ for minimization ($\mu \gg f(x^*) + \delta$ for maximization), that is, when the GP improves the current best value; while the sensitivity to uncertainty is larger for $\mu = f(x^*) \pm \delta$.

Each iteration of active learning proposes a large number of candidate points, and the one with best EI is selected to be added to the dataset. The GP is then re-trained including the new point. At first it is better to prioritize exploration, and as the learning progresses, lean more into exploitation.

The following code shows how to produce a training dataset with active learning using `gpflow` [6,7] to implement GPs, and latin hypercubes to propose the candidates:

Code snippet 2.1: Active Learning

```python
from scipy.stats import norm, qmc
import numpy as np
import gpflow

lhs = qmc.LatinHypercube(d=n_pars)

# Initial points
X = lhs.random(n=N_initial) * (max_X-min_X) + min_X
y = [log_likelihood(x) for x in X]
delta = np.std(y)

for i in range(N_active_learning):
    # Train the GP
    gp = gpflow.models.GPR(
        data=(X, y[:, None]),
        kernel=gpflow.kernels.SquaredExponential(),
        mean_function=None
        )
    opt = gpflow.optimizers.Scipy()
    opt.minimize(gp.training_loss, gp.trainable_variables)

    # Propose candidates and compute EI
```

```
    X_hyp = lhs.random(n=N_candidates)
    X_candidates = X_hyp * (max_X-min_X) + min_X
    mu, sigma = gp.predict_y(X_candidates)
    mu = mu.numpy().flatten()
    sigma = sigma.numpy().flatten()
    I = (mu - np.max(y) - delta)/sigma
    ei = sigma * (I * norm.cdf(I) +  norm.pdf(I))

    # Select best point and add it
    best_idx = np.argmax(ei)
    x_next = X_candidates[best_idx]
    y_next = log_likelihood(x_next)
    X = np.vstack([X, x_next])
    y = np.append(y, y_next)

    # Increase exploitation
    delta *= 0.9
```

As a way to "nudge" the learning algorithm to the correct region of the parameter space, we can add to the pool of candidates some points that we can expect to lie near the best region. For example, one can achieve this by generating random points in the convex hull of the best points.

Code snippet 2.2: Random points in the convex hull

```
from scipy.stats import dirichlet

def points_in_hull(X, y, n_points):
    #select best points
    idx_qtile = np.where(y < np.quantile(y, 0.10))[0]
    # Dirichlet random variable: the sum of all components is 1.
    weights = dirichlet.rvs(
        alpha=np.full(len(idx_qtile), 1/len(idx_qtile)),
        size=n_points
    )

    # Convex combinations of best points
    return weights @ X[idx_qtile]
```

## 2.2 Boosted Decision Trees

The Machine Learning tool that we will use for our analysis is a model able to approximate any arbitrary function $f : \mathbb{R}^n \to \mathbb{R}$, that we will use to create an approximation, or surrogate, of the log-likelihood function of our fit. We have chosen an ensemble method based on regression trees, which is implemented by `xgboost` [8].

Regression trees are a type of decision tree. A decision tree is a diagram that recursively partitions data into subsets, based on the binary (true/false) conditions located at the nodes of the tree. The final subsets in which the data are classified are called "leaves". A decision tree with $T$ leaves is formally a function $q : \mathbb{R}^n \to \{1, 2, \dots, T\}$ which associates to each data point $x \in \mathbb{R}^n$ its leaf $q(x)$. A regression tree assigns to each leaf $i$ a real number $w_i \in \mathbb{R}$. The

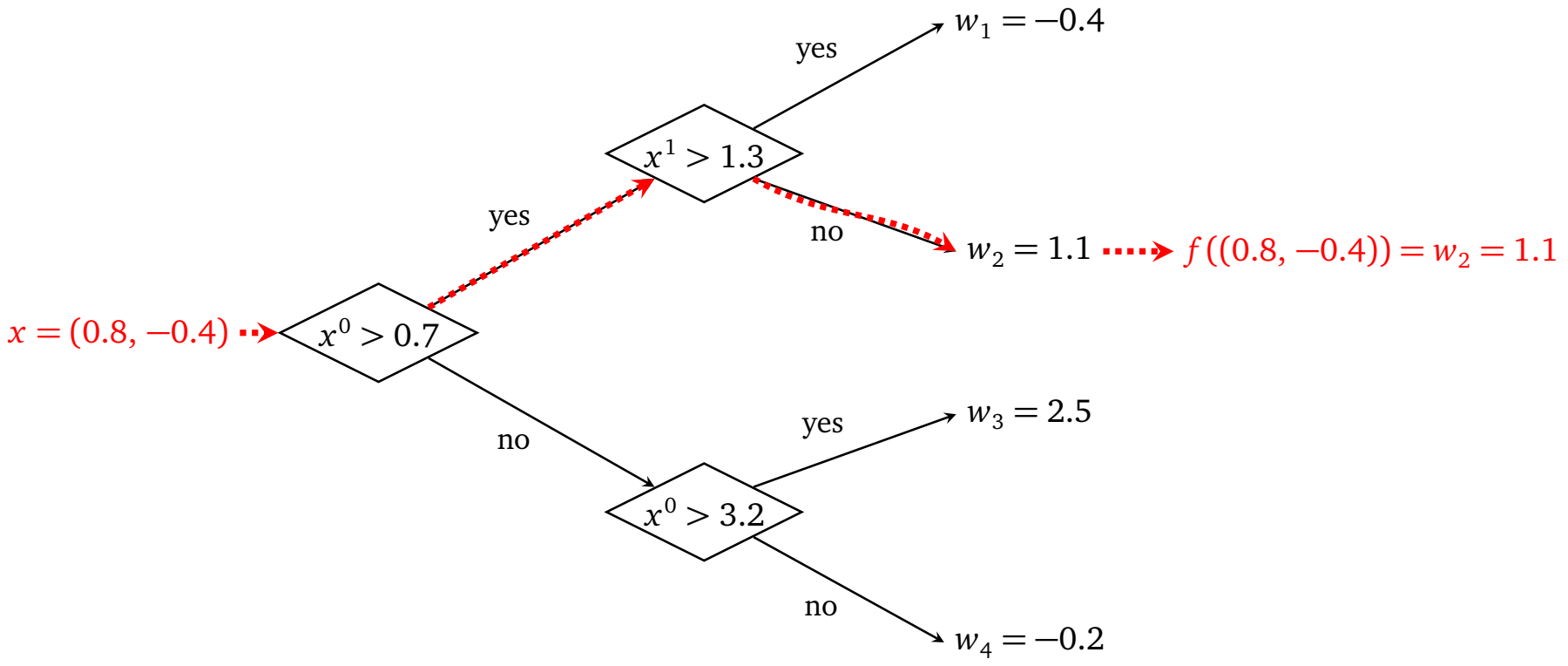

Figure 2: Example of regression tree with four leaves. In red, application of the function $f(x)$ associated with the tree to an input $x$.

regression tree therefore defines a function $f : \mathbb{R}^n \to \mathbb{R}$, given by

$$f(x) = w_{q(x)} . \tag{13}$$

An example of a regression tree with four leaves is depicted in Fig. 2. In practice, a single tree is not general enough to reproduce an arbitrary function, they are "weak learners". For this reason, we consider instead an ensemble of $K$ regression trees $\mathcal{F} = \{f^{(1)}, f^{(2)}, \ldots, f^{(K)}\}$. The ensemble defines a function $\phi : \mathbb{R}^n \to \mathbb{R}$,

$$\phi(x) = \sum_{i=1}^{K} f^{(i)}(x) = \sum_{i=1}^{K} w^{(i)}_{q(x)} . \tag{14}$$

The function $\phi(x)$ will represent the approximation for the log-likelihood function. It will be calculated using supervised learning, that is, the trees are obtained from a dataset $\mathcal{D} = \{(x_i, y_i)\}$ where $x_1, \ldots x_N \in \mathbb{R}^n$ are the inputs and $y_1, \ldots y_N \in \mathbb{R}$ are the pre-computed outputs for each input. In our case, the input data will be the parameters of our model, and the outputs will be the log-likelihood or the $\chi^2$.

In order to train the model from the dataset, we need to define an objective function $\mathcal{L}[\phi]$ that measures how well the model fits the data,

$$\mathcal{L}[\phi] = \sum_i l(\phi(x_i), y_i) + \sum_k \Omega(f^{(k)}) , \tag{15}$$

which has two components:

- The loss function $l(\phi(x_i), y_i)$ is a differentiable function that measures the similarity between the true output $y_i$ and its approximation $\phi(x_i)$. We use as loss function the mean absolute error, $l(\phi(x_i), y_i) = |\phi(x_i) - y_i|$.
- The function $\Omega$ is the regularization term, that penalizes the complexity of trees, that is, trees with many leaves or with large $||w||$. The purpose of the regularization is to prevent overfitting, that is, the model learning "by heart" the training data and being unable to extrapolate from them. Fig. 3 shows the difference between correct convergence and overfitting.

The ensemble is constructed in an iterative way, starting from one single tree $f^{(0)}$ that captures the overall shape of the function, and subsequent trees offer incremental improvements. In order to prevent over-fitting, the shrinkage technique is used, that scales newly added weights by a factor $\eta < 1$, similar to the learning rate in other Machine Learning algorithms.

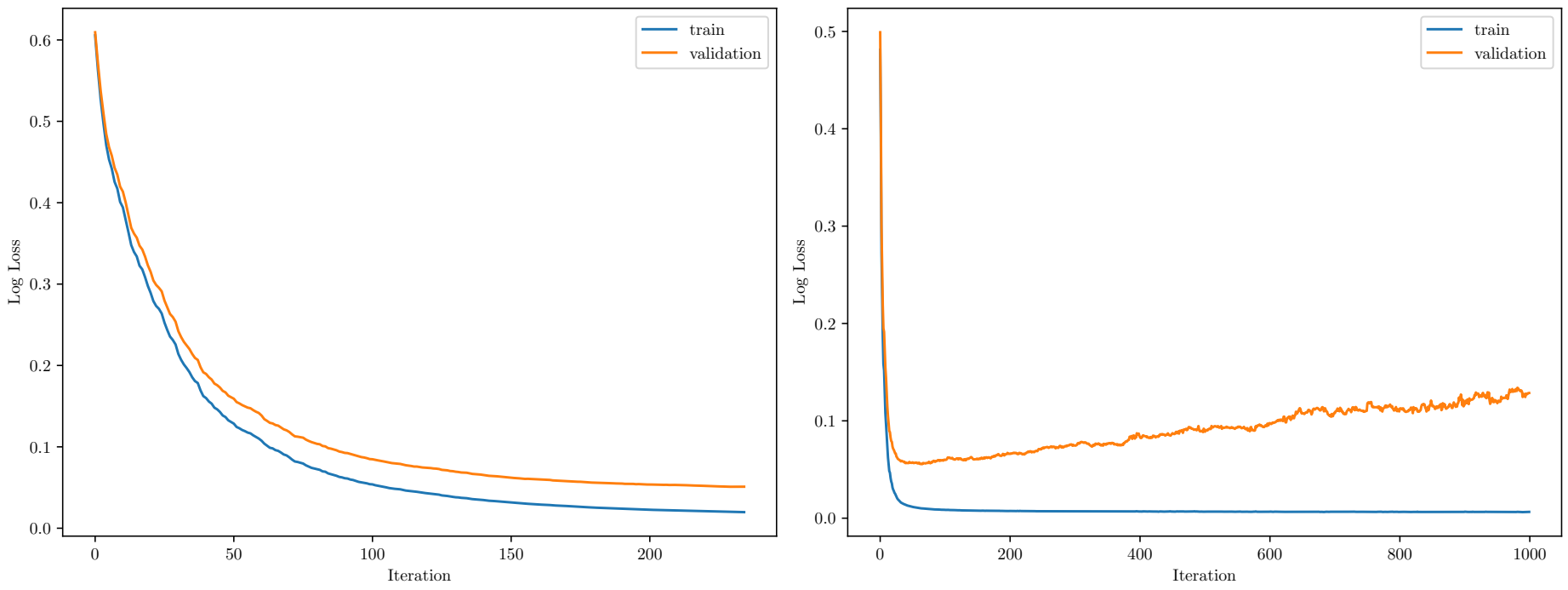

Figure 3: Training evolution of two `xgboost` models for classification tasks. In the model in the left, the loss function in the validation dataset improves at a similar rate as the training dataset, indicating that the model is able to correctly generalize. In the right, with different hyperparameters and without early stopping, the model just learns the training dataset and is not able to generalize to the validation dataset.

### 2.2.1 XGBoost hyperparameters

A hyperparameter is a parameter that controls how the model itself is trained. Some of the most important hyperparameters in XGBoost are:

- **Number of estimators:** Number of trees contained in the ensemble. Typically of order of a few hundreds.
- **Maximum depth:** Maximum number of levels that a tree can have. Large values are more likely to overfit. Typical values are between 5 and 10.
- **Learning rate ($\eta$):** Controls the weight of each tree in the ensemble. Small values reduce overfitting. Typical values are between 0.01 and 0.3.
- **Minimum split loss ($\gamma$):** Minimum loss reduction needed to create a partition on a leaf node. Larger values lead to more conservative algorithms.
- **Regularization parameters ($\alpha$ and $\lambda$):** Penalize large weights, again preventing overfitting.
- **Sampling parameters (subsample, colsample_bytree, colsample_bylevel, colsample_bynode):** Proportion of the training data used to train. Subsample controls the proportion of rows used, while the colsample parameters controls the proportion of columns.

Another useful way to deal with overfitting is by using early stopping: during the training process, the loss function is monitorized in the validation dataset. If it stagnates during a certain number of consecutive training rounds, the process is stopped. This results in an ensemble with less trees than requested by the hyperparameters, since any additional tree would just try to overfit the training dataset.

The following code shows how to train a `xgboost` model from a datset stored in a CSV file:

Code snippet 2.3: Training a xgboost model

```
import xgboost as xgb
import pandas as pd
from sklearn.model_selection import train_test_split
```

```python
import numpy as np

# Load data
df = pd.read_csv("train_data.csv")
pars = ["p1", "p2", "p3"]
X = df[pars]
y = df["y"]
# Create training and validation datasets
X_tr, X_v, y_tr, y_v = train_test_split(X, y, test_size=0.2)
d_tr = xgb.DMatrix(
    X_tr.values.astype(np.float32),
    label=y_tr.values.astype(np.float32)
)
d_v = xgb.DMatrix(
    X_v.values.astype(np.float32),
    label=y_v.values.astype(np.float32)
)
# Train the model
xgb_params = {
    'objective': 'reg:squarederror',
    'tree_method': 'hist',
    'max_depth': 6,
    'learning_rate': 0.1,
    'gamma': 0,
    'subsample': 0.8,
    'colsample_bytree': 0.8,
    'eval_metric': 'mae',
    'nthread': -1, # Use all available CPU cores
}
evals = [(d_tr, 'train'), (d_v, 'eval')]
model_regr = xgb.train(
    xgb_params,
    d_tr,
    num_boost_round=5000,
    evals=evals,
    early_stopping_rounds=5,
    verbose_eval=False
)
# Wrap prediction in a friendlier format
predict_point = lambda x: model_regr.predict(xgb.DMatrix([x,]))
```

It is important to check how the training process progresses in both the training and validation datasets. `xgboost` offers several evaluation metrics: in regression problems, the common choices are mean absolute error (`mae`) and root mean squared error (`rmse`), which penalizes more heavily outliers. Let us modify the previous code to keep track of both of them:

Code snippet 2.4: Evaluation metrics in xgboost

```python
import matplotlib.pyplot as plt
```

```
xgb_params |= {'eval_metric': ['rmse', 'mae']}
eval_results = {}
model_regr = xgb.train(
    xgb_params,
    d_tr,
    num_boost_round=5000,
    evals=evals,
    evals_result=eval_results
    early_stopping_rounds=5,
    verbose_eval=False
)
for metric in ['rmse', 'mae']:
    plt.plot(eval_results['train'][metric], label='train')
    plt.plot(eval_results['eval'][metric], label='validation')
    plt.xlabel('Iteration')
    plt.ylabel(metric)
    plt.legend()
    plt.show()
```

Note that early stopping is decided based on the last metric in `eval_metric`, evaluated at the last dataset in `evals`.

The choice of hyperparameters sometimes can be an artisanal process. However, there are automated tools to tune and evaluate hyperparameters, like cross-validation, and implementations like the `model_selection` module of `sklearn` [9], and `optuna` [10]. In the following example, we use `optuna` to choose the regularization hyperparameters:

Code snippet 2.5: Hyperparameter tuning with optuna

```
import optuna
from sklearn.metrics import mean_absolute_error

def optuna_obj(trial):
    optuna_params = {
        'reg_alpha': trial.suggest_float('reg_alpha', 0, 1.2),
        'reg_lambda': trial.suggest_float('reg_lambda', 0, 1.2),
    }
    params = xgb_params | optuna_params

    model_regr = xgb.train(
        params,
        d_tr,
        num_boost_round=5000,
        evals=evals,
        early_stopping_rounds=5,
        verbose_eval=False
    )
    predictions = model_regr.predict(d_v)
    return mean_absolute_error(y_v, predictions)

optuna.logging.set_verbosity(optuna.logging.WARNING)
```

```python
study = optuna.create_study(direction='minimize')
study.optimize(optuna_obj, n_trials=100, n_jobs=-1)

xgb_params |= study.best_params
```

### 2.2.2 Storing and re-using models

Once a model has been trained, it can be saved into a text-based JSON file (`.json`) or a binary file of type UBJSON (`.ubj`). Binary files have the advantage of smaller file sizes and faster processing times.

Alternatively, using the library `tl2cgen`, XGBoost and other decision tree models can be compiled into C libraries. In addition to storing the trained model, this results in an important speed-up in its evaluation.

The following example stores a `xgboost` model to UBJSON file and compiles it in a Linux system (other operating systems might need different toolchains and library extensions):

Code snippet 2.6: Storing and compiling `xgboost` models

```python
import tl2cgen
import treelite.frontend
# Save to UBJSON
model_regr.save_model('xgb_regressor.ubj')
# Compile to C library
tree_reg = treelite.frontend.from_xgboost(model_regr)
tl2cgen.export_lib(
    tree_reg,
    toolchain='gcc',
    libpath='xgb_regressor.so',
    params={'parallel_comp': 4}
)
```

## 2.3 Explainable models: SHAP values

We can asses the importance of each parameter in the Machine Learning surrogate at any point of the generated samples by using SHAP values [11,12]. SHAP values are based in Lloyd Shapley's work on game theory [13], who won the Nobel Prize in Economics for it in 2012.

The SHAP values are designed with three properties in mind:

- **Local accuracy:** The sum of the SHAP values is equal to the model prediction, at any point of the parameter space.
- **Missingness:** If any feature is missing, its SHAP value is zero.
- **Consistency:** If the model is changed so any feature has larger impact, its SHAP value will increase.

It is important to note that SHAP values provide interpretability at the local level: the prediction of each parameter point is explained as a sum of contributions coming from each parameter. Obviously, by computing the SHAP values in the whole dataset, we also gain global interpretability, as we learn which parameters are more important over all the parameter space.

Given a model $\phi(x)$, the SHAP trains $2^n$ new models $\phi_z(x)$ for $z \in \{0,1\}^n$ binary vectors. The model $\phi_z(x)$ contains the feature $x^{(i)}$ only if $z^{(\alpha)} = 1$, while that feature is ignored when training if $z^{(\alpha)} = 0$. The marginal contribution $\phi_{z'}(x_i) - \phi_z(x_i)$ for two models differing only in the presence of one feature (i.e. $z^{(\alpha)} = 0$, $z'^{(\alpha)} = 1$ and $z^{(\beta)} = z'^{(\beta)}\ \forall\ \beta \neq \alpha$), gives the importance of adding the feature $\alpha$ to the model $z$. The SHAP value for the feature $\alpha$ in the point $x_i$ is just the weighted average of all marginal contributions, with the weight given by a combinatorial factor. An example is depicted in Fig. 4. The prediction without any features $\phi_{0\cdots 0}$ is simply the average of the values $y_i$ in the dataset, and acts as a base value common for all $x_i$.

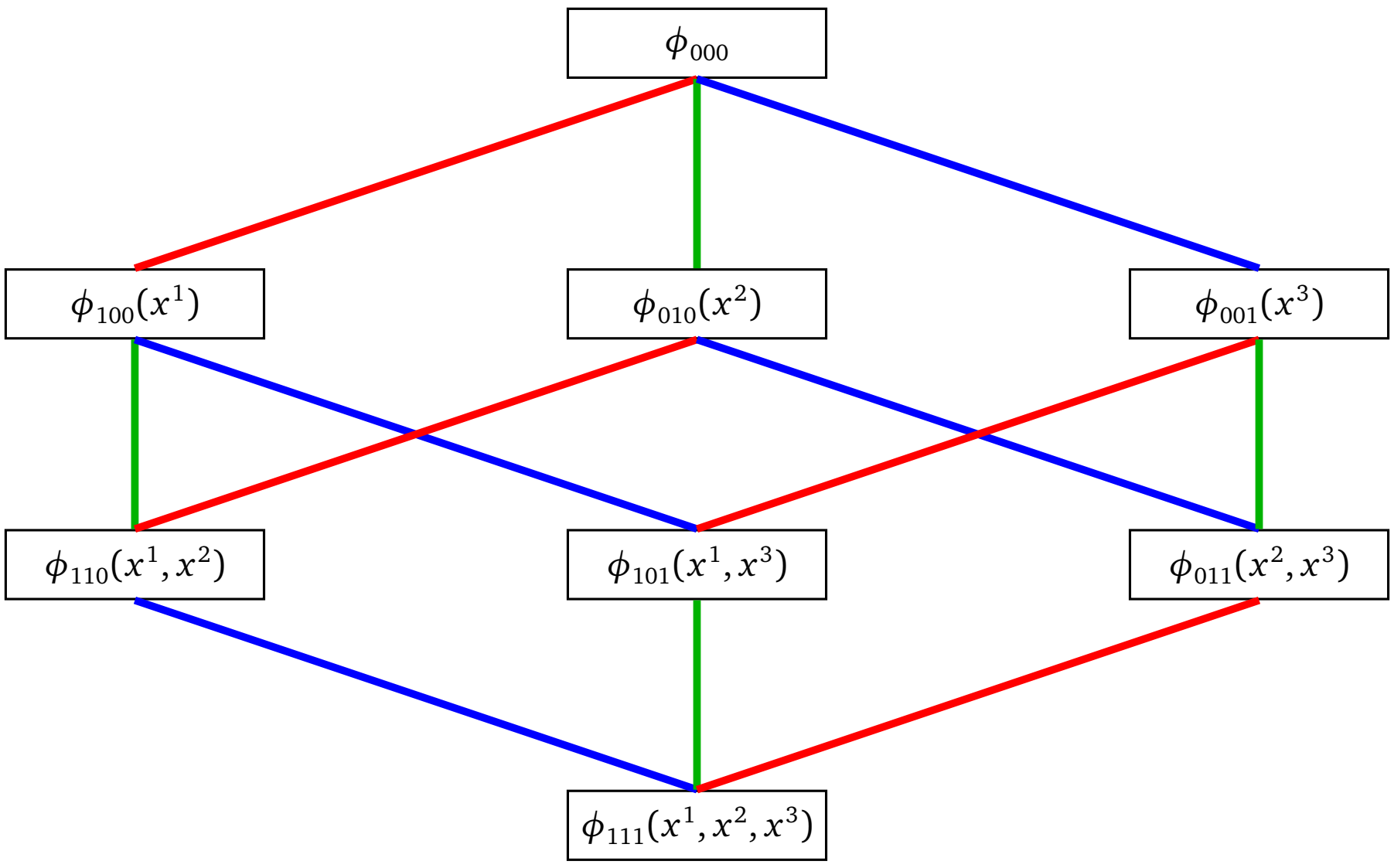

Figure 4: Prediction models that would we necessary to train for three features in order to calculate the SHAP values. The edges represent the marginal contributions for each feature: in red for $x^1$, green for $x^2$ and blue for $x^3$.

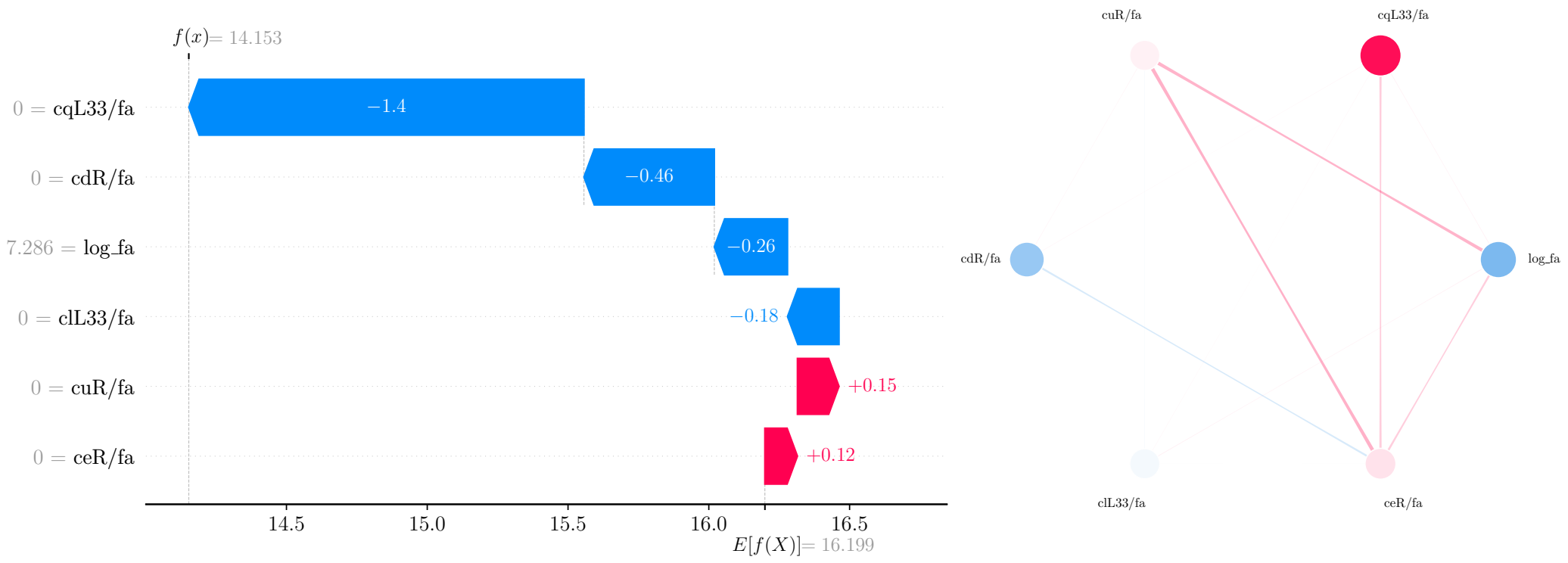

Figure 5: Local explanations of ML predictions: (Left) SHAP values arranged in a waterfall plot, (Right) `shapiq` interaction values arranged in a network plot.

In a generic ML model, training $2^n$ models becomes prohibitive even for moderate $n$. Luckily, in the case of decision trees and tree ensembles, there is a very efficient polynomial algorithm called TreeSHAP [12]. TreeSHAP does not need to train additional models, instead, it can obtain the SHAP values from the model with all features present just by travel all possible paths and keeping track of the contribution of each parameter to the branch splittings.

The following code retrieves the contribution of each parameter to the prediction of one point `x0`, as shown in Fig. 5(Left):

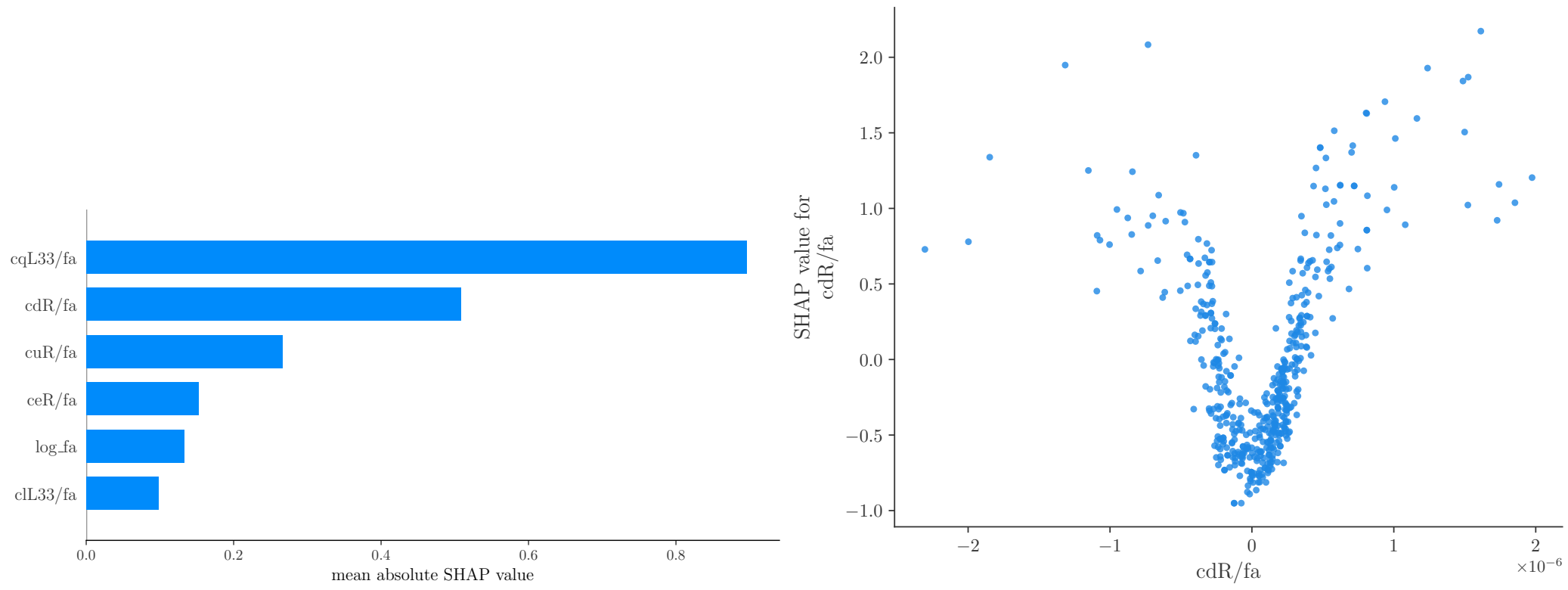

Figure 6: Global explanations of ML predictions: (Left) Averages of SHAP values, (Right) Dependence of SHAP values on the parameter.

Code snippet 2.7: shap for local explainability

```
import shap

explainer = shap.TreeExplainer(model_regr)
print(explainer(x0))
shap.waterfall_plot(explainer(x0)[0])
```

For global explainability, if we have a dataset of points, we can obtain the average contribution of each parameter to the predictions, as shown in Fig. 6(Left), using

Code snippet 2.8: shap for global explainability

```
shap_values = explainer.shap_values(dataset)
shap.summary_plot(shap_values, dataset, plot_type='bar')
```

We can also visualize how the values of the parameter impact the model prediction with

Code snippet 2.9: shap dependence plots

```
shap.dependence_plot(parameter, shap_values, dataset)
```

If the SHAP values for a given parameter show little vertical spread, this is an indication that the prediction shows a strong functional dependence on that parameter. If the SHAP values have large spreads, however, this points to interactions between various parameters.

A generalization of SHAP values to investigate the impact of interaction is implemented by the library `shapiq` [14]. Part of the importance of each parameter is explained instead as importance of pairs, triplets, etc of parameters.

The following code calculates the importance of single parameters and pairs of parameters in explaining the prediction of one point:

Code snippet 2.10: Feature interactions with shapiq

```
import shapiq

explainer_iq = shapiq.TreeExplainer(model_regr, max_order=2)
inter_values = explainer_iq.explain(x0)
print(inter_values)
```

```python
shapiq.network_plot(inter_values)
```

## 2.4 Sampling the posterior distribution

Once we have created an useful approximation of the log-likelihood of interest, let us put it into use. The natural next step is to explore the parameter space by examining the points distributed according to the probability distribution

$$p(\theta|\text{data})\,. \tag{16}$$

This is known as the **posterior distribution**, and it is related to the likelihood function through Bayes' theorem

$$p(\theta|\text{data}) = \frac{p(\theta)}{p(\text{data})} p(\text{data}|\theta)\,. \tag{17}$$

In this expression, $p(\text{data})$ is a normalization constant, that in most sampling algorithm is not even needed. Then, the only missing ingredient is $p(\theta)$, the **prior distribution**. The priors encapsulate our knowledge (or lack thereof) of the model parameters before confronting with the experimental data. An example of informative prior could be a gaussian distribution centered around $\theta_0$ and covariance matrix $C_\theta$,

$$\log p(\theta) = -\frac{1}{2}(\theta - \theta_0)^T C_\theta^{-1}(\theta - \theta_0) + K\,, \tag{18}$$

while for uninformative priors, we can consider a uniform distribution in the support of the parameters,

$$\log p(\theta) = \begin{cases} K & a \leq \theta \leq b\,, \\ -\infty & \text{else}\,. \end{cases} \tag{19}$$

In practice, strong boundaries like in the uniform distribution might result in the sampling algorithm not properly exploring the parameter space, specially in narrow topographies. A possible improvement is to soften the boundaries, e.g. using a sigmoid,

$$\log p(\theta) = -\log\left(1 + \exp\frac{a-\theta}{w}\right) - \log\left(1 + \exp\frac{\theta - b}{w}\right), \tag{20}$$

where $w$ controls the depth fo the wall.

Markov chain Monte Carlo (MCMC) is perhaps the most popular algorithm to sample from an arbitrary probability distribution. A Markov chain is a process where the probability to transition from one state (in our case, a point in parameter space) to the next depends only on the state, and not on the previous history. At each iteration, the MCMC algorithm proposes candidate points which are accepted or rejected in a way that the chain asymptotically reproduces the probability distribution.

We will use the MCMC algorithm implemented by the library `emcee` [15], which is the affine-invariant ensemble sampler [16]. It consists of an ensemble of walkers that explore the parameter space one at a time, based in the position of the other walkers, in a way that is invariant under affine transformations of the parameter space. It has the advantage of working well even in non-differentiable log-probabilities (as is the case of a `xgboost` model), but only for moderate number of parameters ($\lesssim 50$). The proposed candidate for the point at the $n$-th step of the walker $p$ is

$$\theta_k^p = \theta_{k-1}^p + \gamma(\theta_{k-1}^q - \theta_{k-1}^p) \tag{21}$$

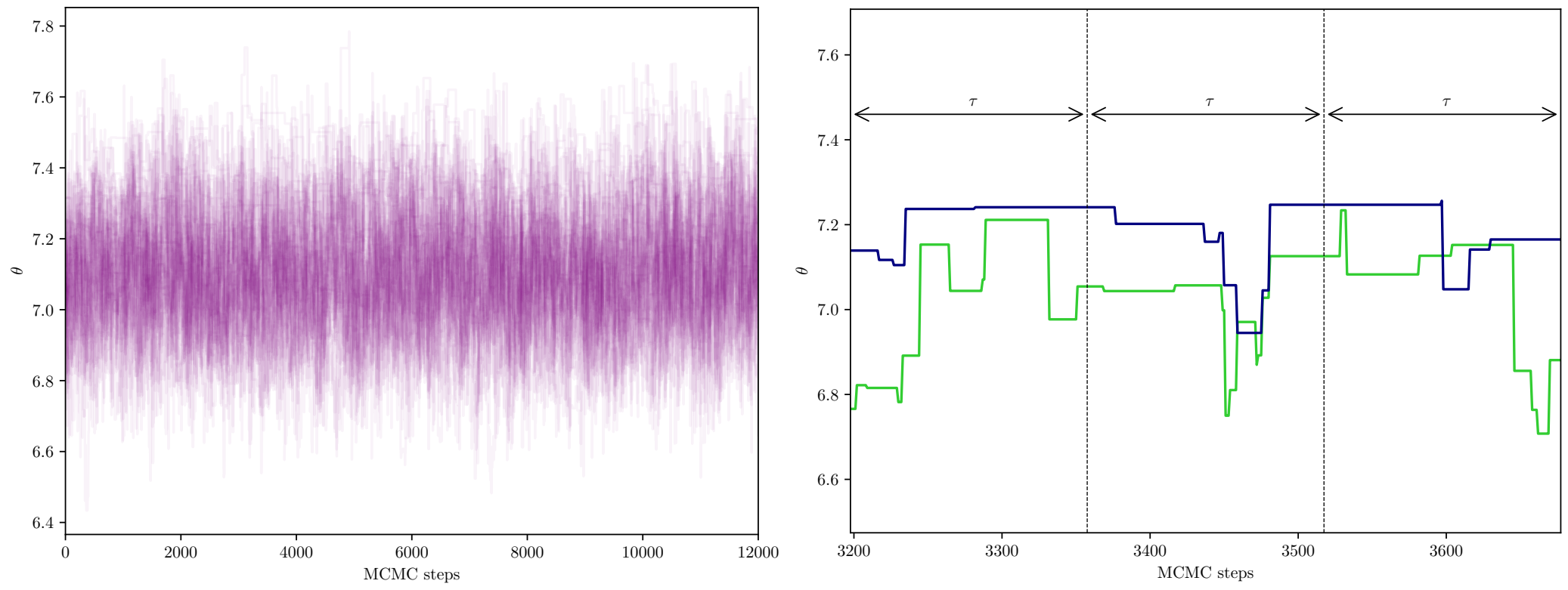

Figure 7: MCMC chains computed with `emcee`: (Left) Values of the parameter $\theta$ in 50 chains from the same ensemble. The densest regions correspond to larger posterior probability. (Right) Detail of two of those chains over 3 autocorrelation times. The parameter remains almost constant during periods of order $\tau$, showing that the points in the chain are correlated.

where the stretch factor $\gamma$ is drawn from the probability distribution

$$p(\gamma) = \begin{cases} \gamma^{-1/2} & a^{-1} \leq \gamma \leq a, \\ 0 & \text{otherwise.} \end{cases} \tag{22}$$

That is, the walker $p$ moves in the direction of the walker $q$, but it might undershoot ($\gamma < 1$) or overshoot ($\gamma > 1$). The candidate move is accepted with transition probability

$$p_T(\theta_k^p | \theta_{k-1}^p) = \min\left[1, \gamma^{d-1} \frac{p(\theta_k^p|\text{data})}{p(\theta_{k-1}^p|\text{data})}\right]. \tag{23}$$

If the candidate move is rejected, $\theta_{k-1}^p$ is added again to the chain.

The basic operation of `emcee` is illustrated in the following code:

Code snippet 2.11: Sampling the posterior with emcee

```
import emcee

n_walkers = 20
def log_posterior(x):
    # chi2 and log_priors defined elsewhere
    return -0.5 * chi2(x) + log_priors(x)

sampler = emcee.EnsembleSampler(
    n_walkers,
    n_pars,
    log_posterior
)

_ = sampler.run_mcmc(
    initial_guess, # list of n_walker guesses, close to minimum
    n_mcmc_steps
)
```

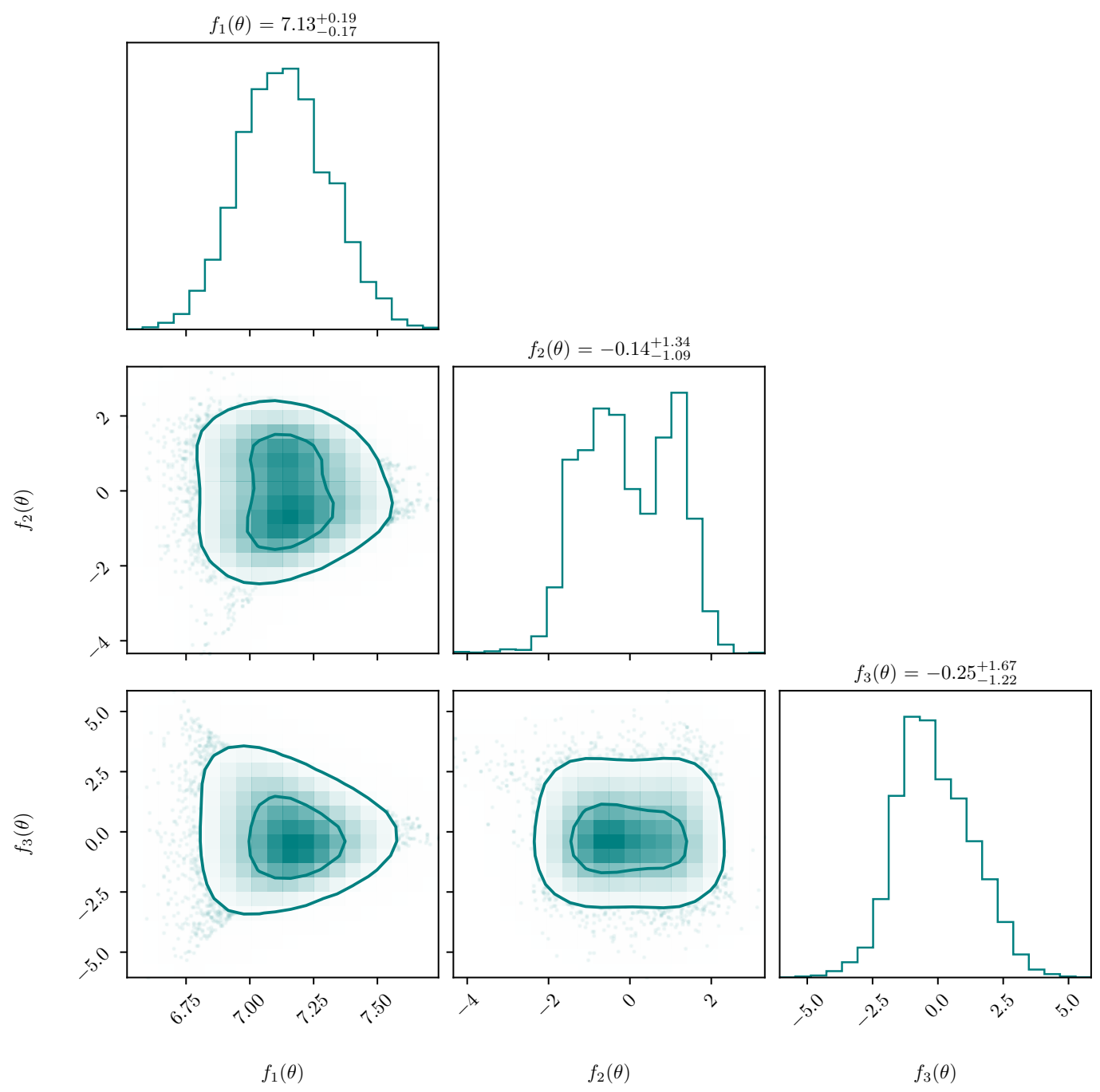

Figure 8: Corner plot of the posterior distribution of three functions of the parameters.

```
print(sampler.acceptance_fraction) # Fraction of points accepted
print(sampler.get_autocorr_time()) # Autocorr. time for each par.
chain = sampler.get_chain(flat=True)
```

One of the drawbacks of MCMC is that the samples are correlated, since they originate from a Markov chain. The integrated correlation time indicates the number of steps in the chain necessary for two points to be statistically independent. These effect can be visualized in Fi. 7 Therefore, the size of the length must be chosen to be much larger that the integrated correlation time, so we are not effectively sampling the same point over and over. This allows us to reduce the number of stored points (thinning) by using just 1 in every $n$ chain. Another related issue is the sensitivity to initial guesses, which can be reduced by discarding (burn-in) the initial points of the chain, typically a few integrated correlation lengths. `emcee` can compute the integrated correlation time of every chain, and offers options for thinning and burn-in.

Once the chain is obtained, it is immediate to calculate the expected value of any function of the parameters under the posterior probability distribution, for example

Code snippet 2.12: Inference with MCMC chains

```
mean_x0 = np.mean(chain[:,0])
variance_x0 = np.mean((chain[:,0] - mean_x0)**2)
```

The library `corner` [17] produces the 1D and 2D confidence plots for parameters or functions of parameters calculated via any MCMC chain

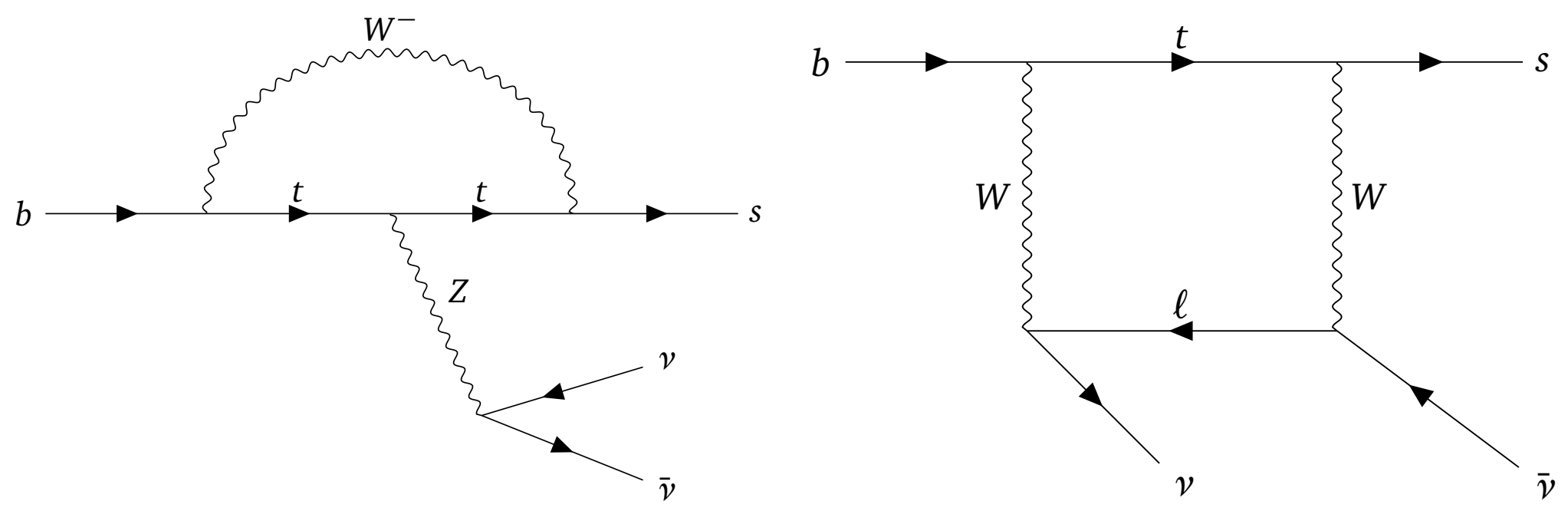

Figure 9: Feynman diagrams for the $b \to s\nu\bar{\nu}$ process in the SM: penguin diagram (left) and box diagram (right).

Code snippet 2.13: corner plots

```
import corner

f1_chain = [f1(x) for x in chain]
f2_chain = [f2(x) for x in chain]
sampled_values = np.vstack([f1_chain, f2_chain]).T

_ = corner.corner(
    sampled_values,
    levels = [0.393, 0.864], # 1sigma and 2sigma in 2D
    smooth = 1.5,
    show_titles = True
)
```

## 3 Application: $B^+ \to K^+\nu\bar{\nu}$ anomaly at Belle II

In the Standard Model (SM), changes of quark flavour are mediated by $W^\pm$ bosons, and therefore occur between quarks of different charges (from up-type to down-type quark and *vice versa*). The transition amplitudes are given by the elements of the CKM matrix, which features a hierarchical structure: transitions between quarks of the same generation are highly favoured over transitions between different generations.

If we turn our attention to hadrons (composite systems of two or more quarks), it is particularly interesting to study decays where a hadron decays into another where only one quark changes flavour without changing electric charges, known as Flavour Changing Neutral Current (FCNC). In the SM, a FCNC hadronic decay must occur with at least two $W^\pm$ emissions, and at least one of them is suppressed because it changes the generation of the quark. Additionally, since the $W^\pm$ appear only in internal lines of the Feynman diagram, there must be at least one closed loop, which supposes another suppression factor. Two examples of topologies of one-loop FCNC decays are shown in Fig. 9, known as penguin and box diagram. All in all, in the SM, FCNC processes are predicted to be extremely rare. On the other hand, a new particle beyond the SM could in principle interact with quarks without respecting the hierarchical structure of the SM, and mediate FCNC even at tree level. In conclusion, FCNC processes are excellent probes for new physics.

Recently, the Belle II experiment studied the FCNC decay $B^\pm \to K^\pm\nu\bar{\nu}$ [18] from a $B^\pm$ meson composed by $bu$ quarks to a kaon $K^\pm$ composed by $su$ quarks, hence a $b \to s\nu\bar{\nu}$ transition.

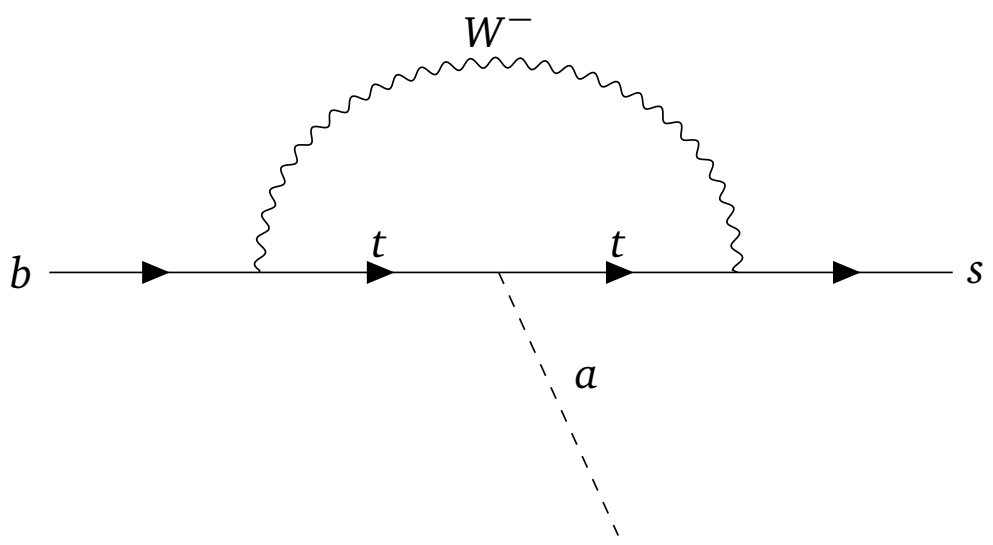

Figure 10: Feynman diagram for the $b \to sa$ process where the effective operator is generated by ALP interactions with $t$ quarks.

It found a branching ratio for this process which is $2.7\,\sigma$ larger than the SM prediction. While this result is not significant enough to claim the discovery of new physics, it has spurred a lot of interest in the phenomenology community. Here we will use the Machine Learning tools to examine one tentative model to explain it, with a light ($m \approx 2\,\mathrm{GeV}$) particle.

## 3.1 Light new physics: Axion-like particles

Axions are very light pseudoscalar particles that were proposed as a solution to the strong-CP problem caused by the fact that the neutron has a negligible dipolar magnetic moment. Axion-like particles (ALPs) are also light (although not necessarily as light) pseudoscalar particles that arise naturally in many extensions of the SM. The most generic interaction of the ALP with SM fermions is given by

$$\mathcal{L} = \frac{\partial_\mu a}{f_a}\left[c_{q_L}^{ij}\bar{q}_i\gamma^\mu q_j + c_{u_R}^{ij}\bar{u}_i\gamma^\mu u_j + c_{d_R}^{ij}\bar{d}_i\gamma^\mu d_j + c_{\ell_R}^{ij}\bar{\ell}_i\gamma^\mu \ell_j + c_{e_R}^{ij}\bar{e}_i\gamma^\mu e_j\right], \tag{24}$$

where $a$ is the ALP, $q_i$ and $\ell_i$ are the $SU(2)$ doublets of left-handed quarks and leptons, and $u_i$, $d_i$ and $e_i$ are the $SU(2)$ singlets of right-handed up-type quarks, down-type quarks and charged leptons. $f_a$ is the decay constant of the ALP (with units of energy): since the operators have dimension 5, the interactions are not renormalizable, and the Lagrangian is an effective Lagrangian valid only up to energies $\Lambda \sim 4\pi f_a$. At higher energies, heavier particles will appear.

As a simplifying assumption, we will demand that the ALP coefficients are flavour-diagonal ($c_f^{ij} = 0$ if $i \neq j$) at the scale $\Lambda$. However, the ALP-fermion couplings depend on the energy scale of the process, governed by the Renormalization Group Equations (RGE). In particular, the RGE evolution will generate flavour-violating effects, such as $b \to sa$ decays (see Fig. 10), which are largely dominated by the coupling to top quarks ($c_t = c_{q_L}^{33} - c_{u_R}^{33}$) [19–25].

Looking at the $B^+ \to K^+\nu\bar{\nu}$ process, we could expect an ALP-mediated decay $B^+ \to K^+ a$ followed by $a \to \nu\bar{\nu}$. However, the probability of an ALP decaying into a pair of fermions $f\bar{f}$ is proportional to $m_f^2$, and given the smallness of the neutrino masses, this would be astronomically improbable. Luckily for us, Belle II does not really detect neutrinos, so the experimental signal is in fact a $B^+$ meson decaying into a $K^+$ and missing energy and momentum. Refs. [26, 27] showed that the Belle II kinematic distributions are in fact compatible with a two-body decay of the $B^+$ meson into a kaon $K^+$ and a particle with mass $2\,\mathrm{GeV}$. By combining the Belle II data with the bounds for $B^0 \to K^{*0}\nu\bar{\nu}$ decay (another $b \to s\nu\bar{\nu}$ process) at the BaBar experiment, we found in Ref. [28] that the particle should be relatively stable, with a proper decay length $c\tau \geq 80\,\mathrm{cm}$, in order to avoid the detection of its decay products in both experiments.

Now the challenging part is to find a model (preferably one that gives us insights about the physics about $\Lambda$) where the rate $b \to sa$ is large enough to generate the Belle II excess, while at the same time the ALP is sufficiently long-lived. In general, for the former we would need a large value of $|c_t|/f_a$ and for the latter a small value of $|c_t|/f_a$, unless there are large cancellations that preclude ALP decays. In Ref. [28] we found one model with such cancellations, based on a previous proposal by [29] (in that case, the cancellations were used in order to avoid constraints from astrophysical processes in the case of a much lighter ALP). The model features non-universal couplings in the third generation of left-handed fermions, $c_{q_L}^{33}$ and $c_{\ell_L}^{33}$, and universal couplings to right-handed fermions $c_{u_R}^{ij} = c_{u_R}\delta^{ij}$, $c_{d_R}^{ij} = c_{d_R}\delta^{ij}$ and $c_{e_R}^{ij} = c_{e_R}\delta^{ij}$. The effective $b \to sa$ is generated by $c_{q_L}^{33} - c_{u_R}$, while the most dangerous decay mode $a \to \mu^+\mu^-$ (and to a lesser extent also decays into hadrons), are suppressed between the tree level $c_{e_R}$ ($c_{u_R}$ and $c_{d_R}$) and the RGE-generated coupling to muons (light quarks). Eventually, all constraints can be satisfied with $f_a \approx 10^7$ GeV and fermion couplings compatible with a non-universal DFSZ model.

In the tutorial, we will explore a similar scenario, but allowing each of the fermion couplings to vary independently.

### 3.1.1 Model parameters and practical concerns

All amplitudes involving one ALP are proportional to $c_f/f_a$. It could be tempting to use only the ratios $c_f/f_a$, and discard altogether the overall scale $f_a$. However, since the matching to the UV theory and running is performed from $\Lambda = 4\pi f_a$, the size of the RGE-generated flavour violating coefficients is controlled by $\log \frac{\Lambda^2}{m_a^2}$. Therefore, it is crucial to keep $f_a$ as an independent parameter, in order to properly capture the $b \to sa$ effects. Based on our previous result, we will select $\log(f_a/1\,\text{GeV}) \in [6,8]$, with an informative prior around $f_a \approx 10^7$ GeV.

In the original model, as in all DFSZ-like UV models, the ALP couplings to fermions are sines or cosines of the orientation of the vacuum values $(v_1, v_2)$ of two Higgs fields. As such, their values are constrained to the interval $c_f \in [-1,1]$. In order to explore more generic scenarios outside the DFSZ-like paradigm, we will allow some more leeway $c_f \in [-3,3]$. Every ALP coupling is allowed to vary independently, and we will assume uniform priors in this region.

In addition to $B \to K^{(*)}\nu\bar{\nu}$, other observables that also constrain the ALP couplings in the region of interest, and that must be included in the global fit analysis, are the following: leptonic and radiative meson decays ($K_S \to \gamma\gamma$, $K_L \to e^+e^-$, $K_L \to \mu^+\mu^-$, $B_s \to \mu^+\mu^-$) and neutral meson oscillation observables ($\epsilon_K$, $\Delta m_K$, $\Delta m_{B^0}$, $\Delta m_{B_s}$), all of them mediated by off-shell ALPs. We use the library ALP-aca (`alpaca-alps`) [30, 31] to compute the $\chi^2$ as a function of the high-energy parameters. ALP-aca already implements the RGE evolution of the parameters and the calculation of all the observables. In particular, the numerical integration of the RGEs is relatively costly, so that the computation of the $\chi^2$ for a single point takes around 10 seconds in an ordinary computer. Therefore, a Machine Learning approach is very advisable in order to speed up the exploration of the parameter space.

Code snippet 3.1: Using ALP-aca

```
import alpaca

# Define couplings at scale Lambda
couplings_high = alpaca.ALPcouplings({
    'cqL': np.diag([0,0,cqL33]),
    'cuR': cuR,
    'cdR': cdR,
    'clL': np.diag([0,0,clL33]),
    'ceR': ceR
```

```
}, basis='derivative_above', scale=4*np.pi*fa)

# Run down to scale ma
couplings_low = couplings_high.match_run(2.0, 'VA_below')

# Compute chi^2
chi2_obj = alpaca.statistics.get_chi2(
    alpaca.sectors.default_sectors['all'],
    ma = 2.0,
    couplings = couplings_low,
    fa = fa
)
chi2 = chi2_obj.chi2_tot()[0]
```

As an additional complication, we observe that there are regions in the proposed parameter space where the $\chi^2$ is much larger than the value at its minimum. If we tried to naïvely train a Machine Learning model on the full $\chi^2$, the model would learn mostly the range of values of $\chi^2$, instead of the fine details of the regions that matter for inference. Instead, we will train a two-stages Machine Learning model: first a classifier will discard the points where the $\chi^2$ is too large (for example, $\Delta\chi^2 > 10$), and then the regressor will only learn the $\chi^2$ for the points that have not been discarded.

### 3.1.2 Tasks for the tutorial

- Try different parameters of the hyperparameters when training the surrogate. Check which choices increase or decrease overfitting and convergence.
- Select subsets of the training data with low and moderate values of $\chi^2$, and calculate SHAP values in them. How does the importance of each parameter change between subsets? Can you find a physical explanation?
- Identify which parameter is less important to define the surrogate model. Train a simplified surrogate model omitting this parameter, and compare the results between both of them.
- Compute expected values for other quantities, for example $c_t = c_{u_R} - c_{q_L}^{33}$, $c_{\mu\mu}(m_a)$, $c_{bs}(m_a)$.

## 4 Conclusion

The transition from traditional numerical fits to Machine Learning-assisted inference represents a significant shift in how we confront complex theoretical models with experimental data. As demonstrated throughout these notes, the primary challenge in modern global fits is no longer just the statistical methodology, but the computational bottleneck created by expensive predictions.

By adopting the workflow presented here, we gain several key advantages:

- **Efficiency:** Active learning reduces the number of evaluations of the likelihood function needed to generate the training data maintaining a balance between exploring the whole parameter space and exploiting the regions of largest likelihood. Using `xgboost` as a surrogate allows us to perform thousands of likelihood evaluations in a fraction of the time required for full numerical integration of RGEs.

- **Interpretability:** Tools like `shap` and `shapiq` move us beyond "black-box" predictions, allowing us to understand exactly which parameters and interactions between them are driving a fit.
- **Robustness:** Combining a two-stage ML approach (classifier and regressor) ensures that the model focuses its attention on the physically relevant regions of the parameter space, preventing the learner from being overwhelmed by high-$\chi^2$ noise.
- **Posterior Inference:** By implementing Markov chain Monte Carlo (MCMC) sampling through tools like `emcee`, we can effectively map the posterior distribution, combining our likelihood with prior knowledge, to derive parameter expectations and visualize multi-dimensional confidence regions.

Other architectures that we have not reviewed in this lecture notes offer distinct advantages depending on the complexity of the parameter space. Neural Networks, for instance, provide highly flexible function approximation, though they often require more extensive hyperparameter tuning and larger training datasets than decision trees. More recently, differentiable likelihood functions implemented in frameworks like `JAX` or `PyTorch` allow for the use of gradient-based optimization and Hamiltonian Monte Carlo. These methods are significantly more scalable than the affine-invariant ensemble sampler in very high-dimensional spaces[2], provided the underlying physics model is itself differentiable. Additionally, Normalizing Flows represent a powerful alternative by directly learning the posterior distribution $p(\theta|\text{data})$ as a transformation of a simple base density, potentially bypassing the need for traditional MCMC chains entirely. However, for the moderate number of parameters ($\lesssim 50$) and potentially non-differentiable boundaries encountered in many New Physics searches, the combination of `xgboost` surrogates and `emcee` remains a highly effective and "explainable" standard.

The $B^\pm \to K^\pm \nu\bar{\nu}$ anomaly at Belle II serves as a perfect testing ground for these methods. Reconciling a 2.7 $\sigma$ excess with the stability requirements of ALPs requires the kind of high-dimensional, efficient exploration that provided by a ML approach. As we look toward the future of particle physics, these explainable and computationally efficient techniques will be essential for identifying the subtle signatures of New Physics hidden within increasingly complex datasets.

## Acknowledgments

I am extremely grateful to the organizers of the 4th COMCHA school on Computing Challenges celebrated in Zaragoza (Spain) from 8th to 15th April 2026, and specially to Siannah Peñaranda.

**Funding information** JA also thanks the support of the Spanish MINECO/FEDER Grants PID2024-160228NB-I00, funded by MCIN/AEI/10.13039/501100011033.

## A Python libraries used in the tutorial

[2] For example, see the very recent global fit to 374 parameters of the SM Effective Field Theory [32] using `jelli`, based on `JAX`.

| Library | Description | Ref. |
|---|---|---|
| `alpaca-ALPs` | ALP phenomenology and $\chi^2$ | [30,31] |
| `corner` | Visualization of Markov chain Monte Carlo samples | [17] |
| `emcee` | Markov chain Monte Carlo sampling | [15] |
| `gpflow` | Gaussian Processes | [6,7] |
| `matplotlib` | Plotting | [33] |
| `numpy` | Array and mathematical operations | [34] |
| `optuna` | Hyperparameter scan | [10] |
| `pandas` | Dataset operations | [35,36] |
| `scikit-learn` | General ML tools | [9] |
| `scipy` | Scientific computing | [37] |
| `shap` | SHAP values | [11,12] |
| `shapiq` | SHAP for interactions of features | [14] |
| `supertree` | Interactive visualization of decision trees | |
| `tl2cgen` | Compilation of `treelite` models to C libraries | |
| `treelite` | Efficient representation of tree-based models | |
| `xgboost` | Boosted decision trees | [8] |